\documentclass[10pt]{revtex4}
\usepackage{amsmath,amssymb}
\usepackage{graphicx,epstopdf}
\usepackage[usenames,dvipsnames,svgnames]{xcolor}
\usepackage{hyperref}
\usepackage{array}
\usepackage{multirow}
\usepackage{blindtext}
\usepackage{relsize}

\begin{document}
	\title{Wormhole solutions in $f(R)$ gravity theory for Chaplygin gas scenario}
	\author{Bikram Ghosh\footnote {bikramghosh13@gmail.com}}
	\author{Saugata Mitra\footnote { saugatamitra20@gmail.com}}
	\affiliation{Department of Mathematics, Ramakrishna Mission Vidyamandira, Howrah-711202, West Bengal, India\\
}

	\begin{abstract}
			The present paper deals with some wormhole solutions which are obtained by taking two different shape functions along with zero  tidal force. For obtaining wormhole solutions, anisotropic fluid and a equation of state $p_t=-\frac{a}{\rho}$ related by Chaplygin gas are considered where $\rho$ is the energy density, $p_t$ is tangential pressure and $a$ is positive constant. Energy conditions are examined for two different models, and it is found that a major energy conditions are satisfied in a region.
	\end{abstract}
    \maketitle

	\section{Introduction}
	A wormhole is an imaginary intuitive concept in general relativity. It acts like a bridge or tunnel \cite{r1}--\cite{r2} connecting two(or more) asymptotic regions or two different universes. The study of wormhole started long back in 1916 by\cite{r3} in analyzing the Schwarzschild solutions. In 1935, Einstein and Rosen\cite{r4} constructed wormhole type solutions considering an elementary article model as bridge connecting two identical sheets. This mathematical representation of space being connected by a wormhole type solution is known as ``Einstein-Rosen bridge''. Wheeler\cite{r5}--\cite{r6} in 1950's considered wormhole as a object of quantum foam connecting different regions of spacetime (at Planck scale). The pioneering work in static wormhole is done by Morris and Thorne\cite{r2}. The static space-time geometry of this hypothetical astrophysics object is sustained by a single fluid violating null energy conditions(NEC)\cite{r7}--\cite{r9}. In the context of asymptotically flat space-time, this violation of NEC is a consequence of the topological censorship. Thus, it is attractive challenge to find realistic matter sources that can support the wormholes. This type of matter is known as exotic matter and they are necessary to form a wormhole in Einstein gravity theory.\\
	In modified gravity theory, however the violation of energy conditions is not necessary. Recently, many astrophysical observations Einstein-Gauss-Bonnet theory\cite{r10.0}--\cite{r14}, Brans-Dicke theory\cite{r15}--\cite{r18} have showed this violation of NEC can be resolved using modified gravity theories. In Brane world gravity\cite{r19}, many researchers\cite{r20}--\cite{r24} have been found wormhole solutions under different considerations. In Rastall gravity theory, non-violation of null and weak energy conditions are also observed\cite{r25}, null and weak energy conditions are satisfied in $f(R,T)$ gravity theory by choosing suitable form of $f(R,T)$ and other components\cite{r26}--\cite{r27}.\\
	$f(R)$ gravity theory (in which the Lagrangian density is an arbitrary function of the Ricci scalar $R$) is one of the simplest modifications from the geometrical aspect. In $f(R)$ gravity theory, it is observed that by choosing suitable form of $f(R)$ and other comoponents not only null energy conditions satisfied, weak and dominant energy conditions also satisfied\cite{r28}--\cite{r29}. Lobo et al.(2009)\cite{r35} found the wormhole geometries in $f(R)$ gravity and gave some exact solutions by captivating specific shape function and various equation of parameters, and noticed that violation of weak and null energy conditions is not necessary for formation of wormhole solutions in $f(R)$ gravity.
	For obtaining wormhole solutions, many researcher considered Chaplygin gas scenario\cite{r29.1}--\cite{r29.4}.
	\par  In this paper, considering Chaplygin gas scenario $f(R)$ form are evaluated for a given shape function with constant redshift function. In this work, the main motivation is to examine whether the energy conditions are satisfied or not in the obtained wormhole solutions. The remainder of this paper is arranged as follows: in section \ref{sec2}, the necessary field equations on $f(R)$ gravity are constructed. Wormhole solutions are obtained, energy conditions are mentioned and examined also in section \ref{sec3}. In section \ref{sec4}, embedding diagram and proper radial distance are shown. Finally, we discuss our results in section \ref{sec5}.
	\section{Field equations on $f(R)$ gravity}\label{sec2}
	\par In 4-dimensional space-time, the general static and spherically symmetric wormhole is described by the line element \cite{r2}
	\begin{equation}\label{eq1}
	ds^2=-e^{2\phi(r)}dt^2+\frac{dr^2}{1-\frac{b(r)}{r}}+r^2d\Omega^2 , \end{equation}
	where $d\Omega^2=d\theta^2+{\sin^2\theta}d\phi^2$ is the line element on a unit sphere, $\phi(r)$ and $b(r)$ are arbitrary function of radial co-ordinate $`r$', termed as redshift function and shape function respectively. For traversability condition of wormhole, the event horizon should be absent, for which we require $e^{2\phi(r)}\neq0$ {\it i.e.} $\phi(r)$ should be finite everywhere. It is not necessary that $\phi(r)$ is only constant, it could be variable as well. Now for the existence of wormhole, the shape function $b(r)$ is restricted as follows $(i)$ $b(r_0)=r_0$ 
	 ($r=r_0$ is the location of the throat), $(ii)$ $ \frac{b(r)}{r}<1 $ for $r>r_0$, $(iii)$ $\frac{b(r)}{r}\rightarrow0$ as $r\rightarrow\infty$ and $(iv)$ $
	\frac{b-b'r}{2b^2}>0$ is the condition for flare-out\cite{r31}.\\
	 In $f(R)$ gravity model, action is given by \cite{r32}
	\begin{equation}\label{eq2}
	S=\frac{1}{2k^2}\int d^4x\sqrt{-g}f(R)+S_M(g^{\mu\nu},\psi)
	\end{equation} where $k^2=8\pi G$, and for notational simplicity, we consider $k^2=1$ in the present work. $S_M(g^{\mu\nu},\psi)$ is the matter action, defined as $S_M=\int d^4x\sqrt{-g}\mathcal{L}_M(g_{\mu\nu},\psi)$, where $\mathcal{L}_M$ is the matter Lagrangian density, where matter is assumed to be minimally coupled to gravity and $\psi$ collectively denotes the matter fields.\par Now, varying the action with respect to $g^{\mu\nu}$ in the metric approach we get the corresponding field equations\cite{r30}, \cite{r33}-\cite{r34},
	\begin{equation}\label{eq3}
	F(R)R_{\mu\nu}-\frac{1}{2}f(R)g_{\mu\nu}-\nabla_\mu\nabla_\nu F(R)+g_{\mu\nu}\Box F(R)=T^m_{\mu\nu}
	\end{equation}
	where $F=\frac{df}{dR}$ , $\Box =\nabla_\mu\nabla^\mu$, $R_{\mu\nu}$ is the Ricci tensor,
	and $T^m_{\mu\nu}$ is the energy momentum tensor. Considering the trace of the field equations (\ref{eq3}) we obtain the following relation
	\begin{equation} \label{eq4}
	FR-2f+3\Box F=T,
	\end{equation} where $T$ is trace of the energy momentum tensor $T^m_{\mu\nu}$.
	Using equation (\ref{eq4}) in (\ref{eq1}) after some algebraic manipulation one gets the normal form of Einstein field equations as
	\begin{equation}\label{eq5}
	G_{\mu\nu}=R_{\mu\nu}-\frac{1}{2}Rg_{\mu\nu}=T_{\mu\nu}^{\text{eff}}=T_{\mu\nu}^{(g)}+T_{\mu\nu}^m/F,
	\end{equation}
	where $G_{\mu\nu}$ is the Einstein tensor, $T_{\mu\nu}^{\text{eff}}=T_{\mu\nu}^{(g)}+T_{\mu\nu}^m/F$ is the effective energy-momentum tensor and $T_{\mu\nu}^{(g)}$, due to gravity is given by
	\begin{equation}\label{eq6}
	T_{\mu\nu}^{(g)}=\frac{1}{F}\left[\nabla_\mu\nabla_\nu F-\frac{1}{4}g_{\mu\nu}(RF+\Box F+T)\right].
	\end{equation} 
	For obtaining some particular wormhole solution, we assume that the stress-energy momentum tensor that threads the wormhole is given by an anisotropic matter distribution as \cite{r35}
	$$T_{\mu\nu}=(\rho+p_t)u_\mu u_\nu+p_tg_{\mu\nu}+(p_r-p_t)v_\mu v_\nu,$$ where $v^\mu$ is the unit spacelike vector in the radial direction, $u^\mu$ is the four velocity vector, $\rho(r)$ is the energy density, $p_r(r)$ is the radial pressure measured in the direction of $v^\mu$, and $p_t(r)$ is the transverse pressure measured in the orthogonal direction to $v^\mu$.
	\par 
	 We shall take the redshift function to be constant which helps to investigate interesting wormhole solutions.
	 For the metric (\ref{eq1}), the effective field equations can be exposed as\cite{r35},\cite{r41}-\cite{r42}	
\begin{equation}\label{eq13}
\frac{Fb^\prime}{r^2}=\rho,
\end{equation}
\begin{equation}\label{eq14}
-\frac{bF}{r^3}+\frac{F^\prime}{2r^2}(b^\prime r-b)-F^{\prime\prime}\left(1-\frac{b}{r}\right)=p_r,
\end{equation}
\begin{equation}\label{eq15}
-\frac{F^\prime}{r}\left(1-\frac{b}{r}\right)+\frac{F}{2r^3}(b-b^\prime r)=p_t.
\end{equation}
For the curvature scalar, R is given by
	\begin{equation}\label{eq11}
	R=\frac{2b^\prime}{r^2}.
	\end{equation}
	\section{Evaluation of some $f(R)$ and validation of energy conditions}\label{sec3}
	\subsection{$f(R)$ for the different shape function}\label{iiiA}
	\par ~~~~~~For obtaing wormhole solution we consider a equation of state $p_t=-\frac{a}{\rho}$ for Chaplygin gas, where $a$ is positive constant. Using this equation of state, from the field equations(\ref{eq13})--(\ref{eq15}), we obtain
	\begin{equation}\label{ode}
	F(r)F'(r)\frac{b'}{r^3}\left(1-\frac{b}{r}\right)-F^2(r)\frac{b'}{2r^5}(b-b'r)-a=0.
	\end{equation}
	Now for a given $b(r)$, solution of the odinary differential equation (\ref{ode}) is given by
   \begin{equation}\label{eq16}
   F^2(r)\times e^{\int\frac{b-b'r}{r(b-r)}dr}=2a\int\frac{r^4}{b'(r-b)}\times e^{\int\frac{b-b'r}{r(b-r)}dr}dr.
   \end{equation}
	\subsubsection{Evaluation of $f(R)$, shape function $b(r)=\frac{r_0^2}{r}$}
	Let us consider the shape function $b(r)=\frac{r_0^2}{r}$\cite{r36}. For this shape function (1), from the equation (\ref{eq16}) we get the result,
	\begin{equation}\label{eq22}
	F(r)=C_1r_0^2\sqrt{2a\biggl\{\bigg(\frac{r_0}{r}\bigg)^2-1\biggr\}\biggl\{\frac{1}{2}\bigg(\frac{r}{r_0}\bigg)^4+\frac{1}{6}\bigg(\frac{r}{r_0}\bigg)^6+\frac{1}{2}\bigg(\frac{r}{r_0}\bigg)^4\Bigg(1-\bigg(\frac{r}{r_0}\bigg)^2\Bigg)^{-1}+2\bigg(\frac{r}{r_0}\bigg)^2+2ln(r^2-r_0^2)\biggr\}},
	\end{equation}
	where $C_1$ is arbitrary constant.
	In this case Ricci scalar is given by $R=-\frac{2r_0^2}{r^4}$ (using equation (\ref{eq11})) i.e, $r=\left(\frac{-2r_0^2}{R}\right)^{1/4}$, and hence at the throat we have $r_0=\left(\frac{-2}{R_0}\right)^{1/2}$. Putting these relations in equation (\ref{eq22}) gives the form of $F(R)$, which is given by
	\begin{equation}\label{eq23}
	F(R)=-\frac{2C_1}{R_0}\bigg[2a\biggl\{\bigg(\frac{R_0}{R}\bigg)^{-\frac{1}{2}}-1\biggr\}\biggl\{\frac{1}{2}\bigg(\frac{R_0}{R}\bigg)+\frac{1}{6}\bigg(\frac{R_0}{R}\bigg)^{\frac{3}{2}}+\frac{1}{2}\bigg(\frac{R_0}{R}\bigg)\Bigg(1-\bigg(\frac{R_0}{R}\bigg)^{\frac{1}{2}}\Bigg)^{-1}+2\bigg(\frac{R_0}{R}\bigg)^{\frac{1}{2}}+2ln\Bigg(\frac{2}{\sqrt{RR_0}}+\frac{2}{R_0}\Bigg)\biggr\}\bigg]^{\frac{1}{2}}.
\end{equation}
Now we can get the specific form of $f(R)$ from the given integral (continuity of the integrand gives assurance the existence of the integration)
\begin{equation}\label{eq24}
f(R)=\int_{R_0}^{R}F(R)dR ~.
\end{equation}
The above $F(r)$(in equation \ref{eq22}) will be treated as model I.
	
	\subsubsection{ Evaluation of $f(R)$, shape function $b(r)=\frac{r}{1+r-r_0}$, $0<r_0<1$}
	Let us now consider another shape function $b(r)=\frac{r}{1+r-r_0}$, $0<r_0<1$\cite{r36}. For shape function (2), from the equation (\ref{eq16}) we get the result,
	\begin{eqnarray}\label{eq25}
	\nonumber
	F(r)&=&C_2\sqrt{2}\Bigg[\frac{a(r-r_0)}{(1+r-r_0)(1-r_0)}\Biggl\{\frac{r^6}{6}-\frac{2}{5}r^5r_0+\frac{4}{5}r^5+\frac{3}{2}r^4+\frac{1}{4}r^4r_0^2-r^4r_0+\frac{4}{3}r^3+\frac{1}{2}r^2+2r^2r_0+4rr_0^2+2rr_0\\
	&~&-\frac{r_0^3}{r-r_0}
	+3r_0^2ln(r-r_0)+4r_0^3ln(r-r_0)
	\Biggr\}\Bigg]^{\frac{1}{2}},
	\end{eqnarray}
	where $C_2$ is arbitrary constant. Now for this shape function Ricci scalar is given by $R=\frac{2(1-r_0)}{r^2(1+r-r_0)^2}$ (using equation (\ref{eq11})) which gives
	\begin{equation}\label{eq26}
	r=\frac{(r_0-1)\pm\sqrt{(1-r_0)^2+4\sqrt{\frac{2(1-r_0)}{R}}}}{2} 
	\end{equation}
	and at the throat we have 
	$r_0=\frac{-1+\sqrt{1+2R_0}}{R_0}$.
	Since $r$ takes always positive sign so we consider the positive case in equation (\ref{eq26}). 
	 Putting this relations in equation (\ref{eq25}) we obtain the form of $F(R)$ as follows
	 \begin{eqnarray}\nonumber
	 F(R)&=&C_2\sqrt{2}\Bigg[\left(\frac{a}{\alpha}\right)\left(\alpha-2+\sqrt{\alpha^2+4\sqrt{\frac{2\alpha}{R}}}\right)\left(\alpha+\sqrt{\alpha^2+4\sqrt{\frac{2\alpha}{R}}}\right)^{-1}\Bigg\{\frac{1}{3.2^7}\left(-\alpha+\sqrt{\alpha^2+4\sqrt{\frac{2\alpha}{R}}}\right)^6\\\nonumber
	 &~&+\frac{1}{5.2^4}(1+\alpha)\left(-\alpha+\sqrt{\alpha^2+4\sqrt{\frac{2\alpha}{R}}}\right)^5+\frac{(3+2\alpha+\alpha^2)}{2^6}\left(-\alpha+\sqrt{\alpha^2+4\sqrt{\frac{2\alpha}{R}}}\right)^4+\frac{1}{6}\left(-\alpha+\sqrt{\alpha^2+4\sqrt{\frac{2\alpha}{R}}}\right)^3\\\nonumber
	 &~&+\frac{(5-4\alpha)}{2^3}\left(-\alpha+\sqrt{\alpha^2+4\sqrt{\frac{2\alpha}{R}}}\right)^2+(1-\alpha)(3-2\alpha)\left(-\alpha+\sqrt{\alpha^2+4\sqrt{\frac{2\alpha}{R}}}\right)\\
	 &~&-\frac{2(1-\alpha)^3}{\alpha-2+\sqrt{\alpha^2+4\sqrt{\frac{2\alpha}{R}}}}+(7-4\alpha)(1-\alpha)^2\ln\Bigg\{\frac{1}{2}\left(\alpha-2+\sqrt{\alpha^2+4\sqrt{\frac{2\alpha}{R}}}\right)\Bigg\}\Bigg\}\Bigg],	 
	 \end{eqnarray}
	 where $\alpha=\frac{-(1+R_0)+\sqrt{1+2R_0}}{R_0}$. As done in model I, the specific form of $f(R)$ can be obtained by considering the following integral (continuity of the integrand gives assurance the existence of the integration):
	 \begin{equation}\label{int2}
	 f(R)=\int_{R_0}^{R}F(R)dR ~.
	 \end{equation} 
	   The above $F(r)$ (in equation \ref{eq25}) will be treated as model II.
	   \par 
	   It is to be noted that using equation (\ref{eq4}), the form of $f(R)$ can also be computed for the above two models.
	 \subsection{Energy conditions}
	 Let $\scriptsize {S}$ be the four dimensional space-time and $\mathcal{T}_a$ is the tangent space at $a\in S$. Now for any timelike vector $U\in \mathcal{T}_a$, the energy momentum tensor must follow the inequality $T_{\mu\nu} U^\mu U^\nu\geq0$ at every point $a\in S$.
	 The null energy condition (NEC), weak energy condition (WEC), strong energy condition (SEC) and dominant energy condition (DEC) are considered main energy conditions in the background of $f(R)$ gravity. The energy conditions are defined as: NEC$\iff T_{\gamma\xi}W_\mu W_\nu\geq0$, WEC$\iff T_{\gamma\xi}U_\mu U_\nu\geq0$, SEC$\iff \left(T_{\gamma\xi}-\frac{T}{2}g_{\gamma_\xi}\right)U_\mu U_\nu\geq0$, DEC$\iff T_{\gamma\xi}U_\mu U_\nu\geq0$, where $T_{\gamma\xi}U^\gamma$ is not space-like and $g_{\xi\gamma}$ is the metric tensor, $W_\mu$ mentions the null vector and $U_\mu$ represents the timelike vector. 
	 \par The following energy conditions with respect to principal pressures are calculated as:
	 \begin{eqnarray}\nonumber
	 &\boxdot& \text{NEC}\iff \rho+p_j\geq0, \forall j,\\\nonumber
	  &\boxdot& \text{WEC}\iff \rho\geq0 \text{ and } \rho+p_j\geq0, \forall j,\\
	   &\boxdot& \text{SEC}\iff \rho+p_j\geq0,~\rho+\sum p_j, \forall j,\\\nonumber
	    &\boxdot& \text{DEC}\iff \rho\geq0, p_j\in[-\rho, \rho], \forall j.
	 \end{eqnarray}
	 The above conditions can be written as\cite{r28}:
	 \begin{eqnarray}
	 \label{eq13.1} &(I)& \text{NEC} : \rho+p_r\geq0,~ \rho+p_t\geq0,\\\label{eq14.1}
	 &(II)& \text{WEC} : \rho\geq0,~   \rho+p_r\geq0,~  \rho+p_t\geq0,\\\label{eq15.1}
	 &(III)& \text{SEC} : \rho+p_r\geq0,~  \rho+p_t\geq0,~  \rho+p_r+2p_t\geq0,\\\label{eq16.1}
	 &(IV)& \text{DEC} : \rho\geq0,~  \rho-|p_r|\geq0,~  \rho-|p_t|\geq0.
	 \end{eqnarray}\\
	 In section \ref{iiiA} we find two forms of $f(R)$. Now we want to verify how the energy conditions are satisfied. Using field equations (\ref{eq13})-(\ref{eq15}), we examine all the energy conditions for the above two models (Models I and II).
 \begin{figure}[htb!]
 	\centering
 	\begin{minipage}{.42\textwidth}
 		\centering
 		\includegraphics[width=.6\linewidth]{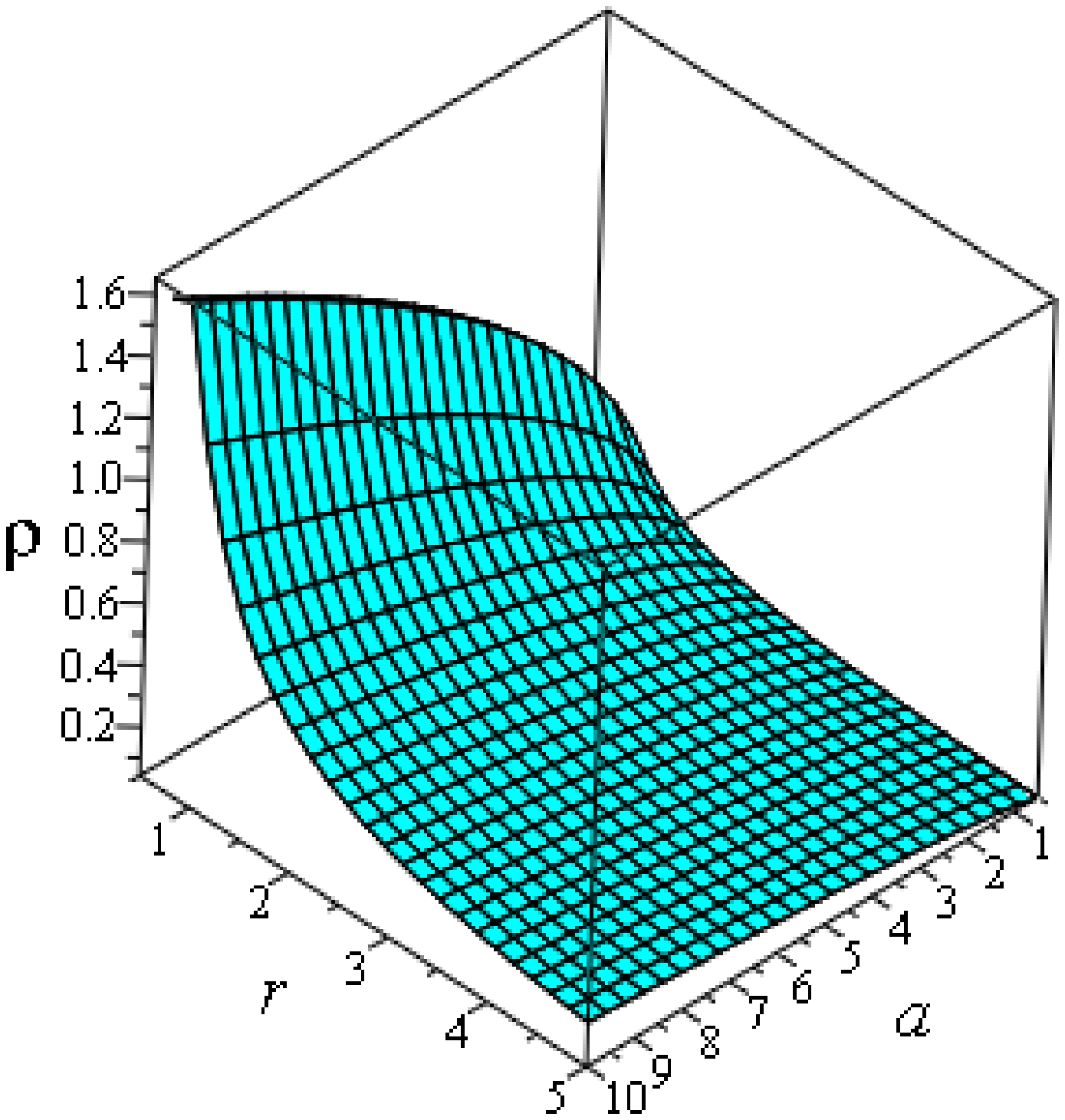}
 		\centering FIG.1(A)
 	\end{minipage}
 	\begin{minipage}{.42\textwidth}
 		\centering
 		\includegraphics[width=.6\linewidth]{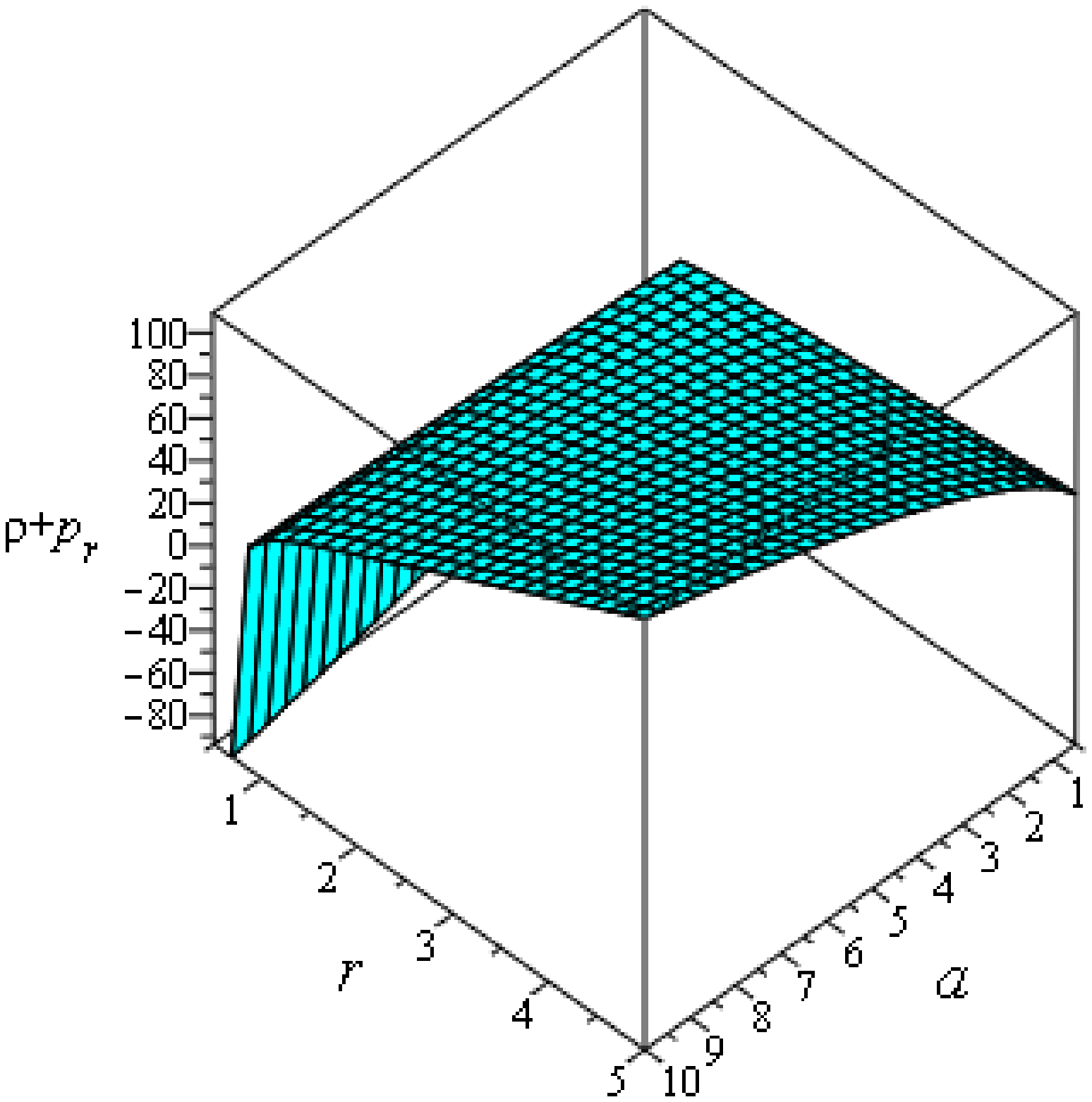}
 		\centering FIG.1(B)
 	\end{minipage}
 	\centering
 	\begin{minipage}{.42\textwidth}
 		\centering
 		\includegraphics[width=.6\linewidth]{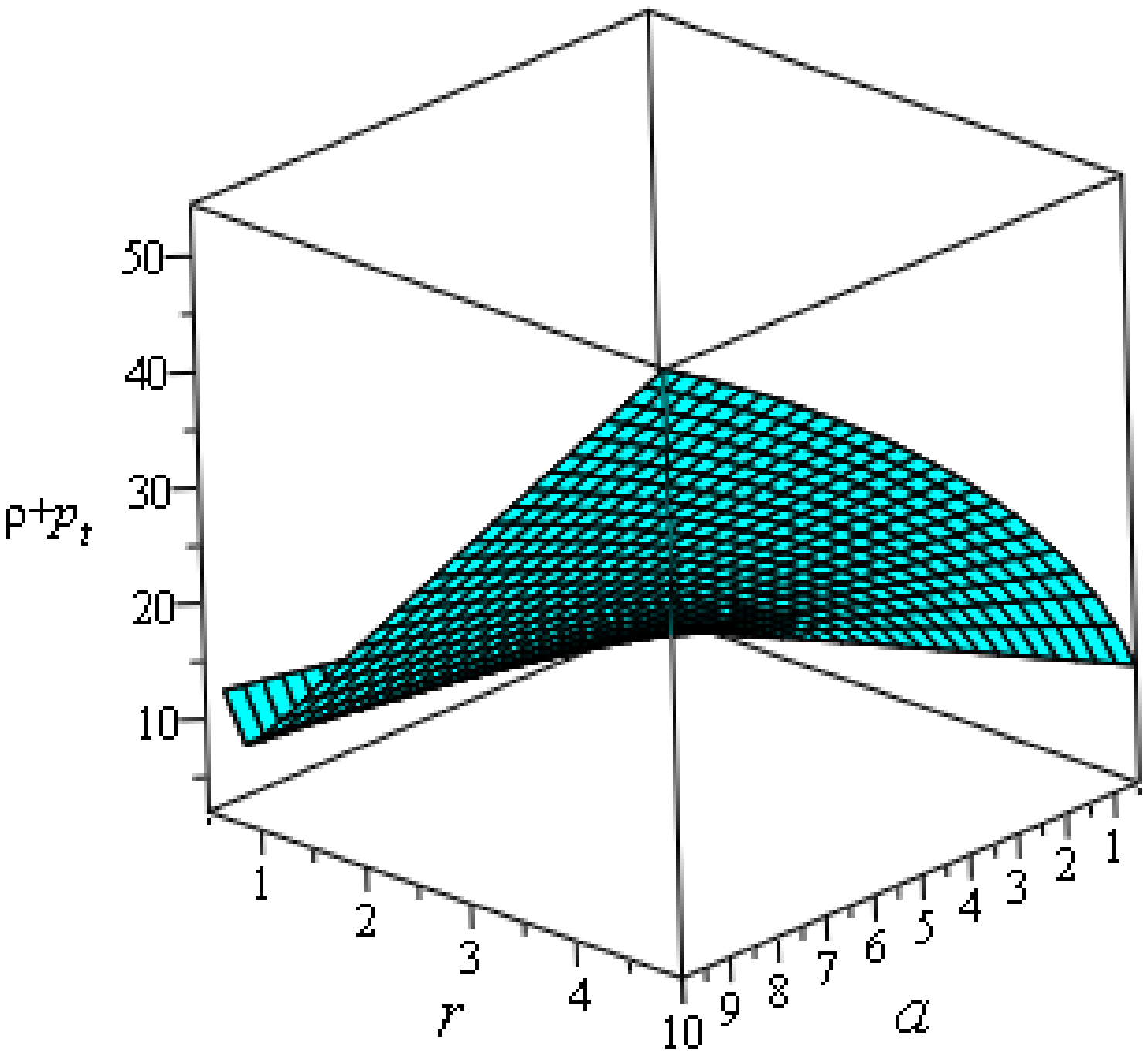}
 		\centering FIG.1(C)
 	\end{minipage}
 	\begin{minipage}{.42\textwidth}
 		\centering
 		\includegraphics[width=.6\linewidth]{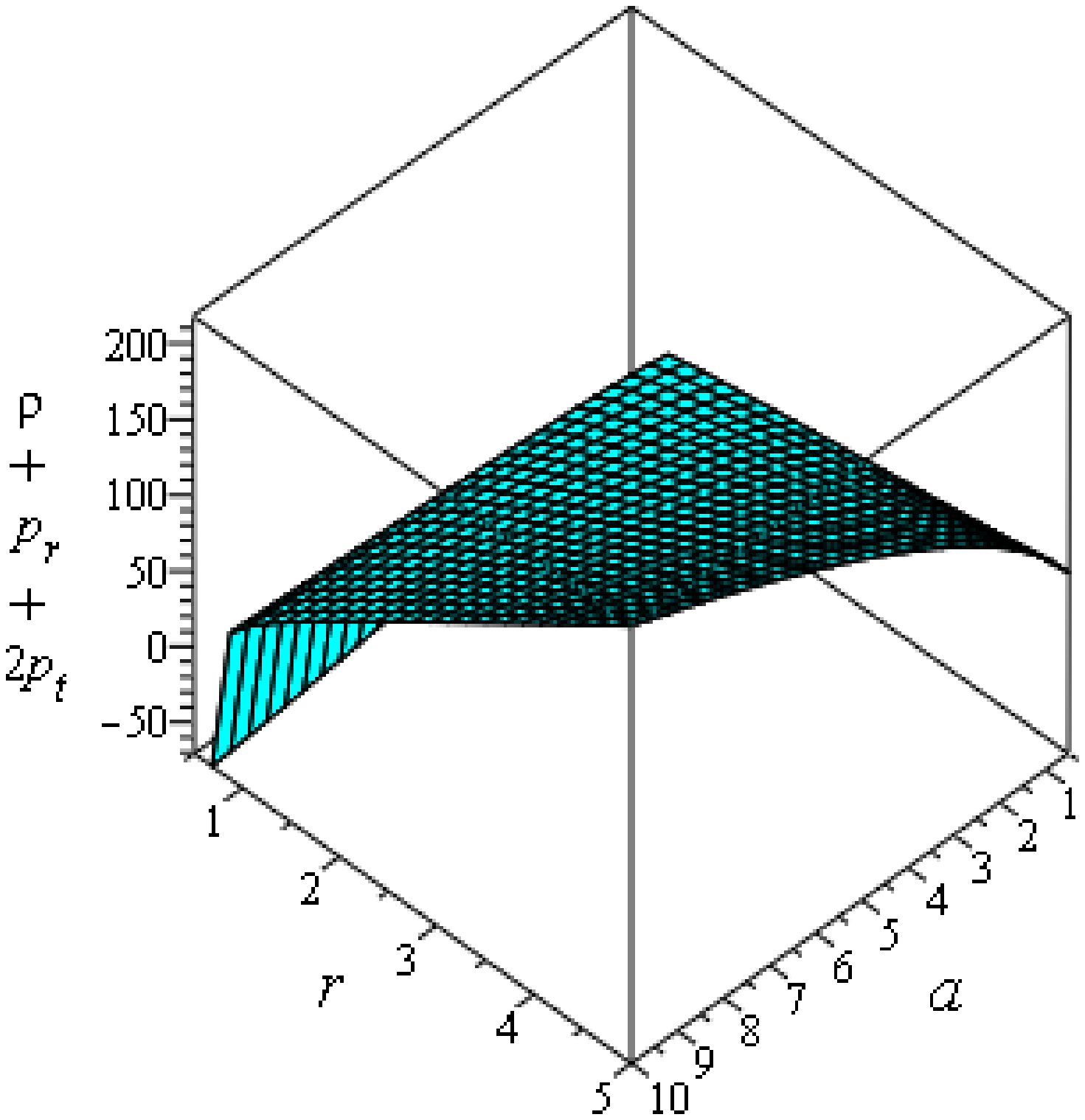}
 		\centering FIG.1(D)
 	\end{minipage}
 	\centering
 	\begin{minipage}{.42\textwidth}
 		\centering
 		\includegraphics[width=.6\linewidth]{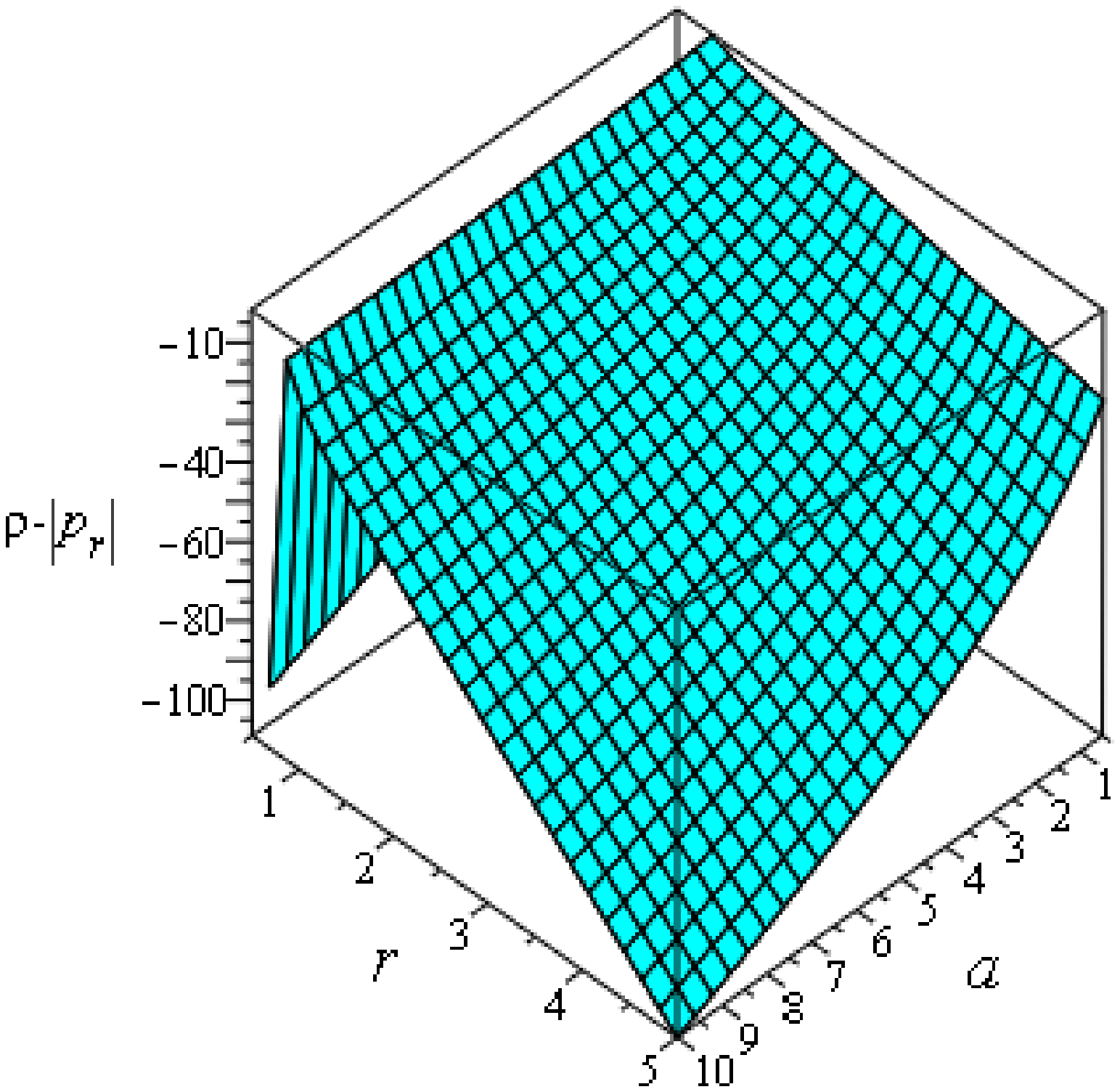}
 		\centering FIG.1(E)
 	\end{minipage}
 	\begin{minipage}{.42\textwidth}
 		\centering
 		\includegraphics[width=.6\linewidth]{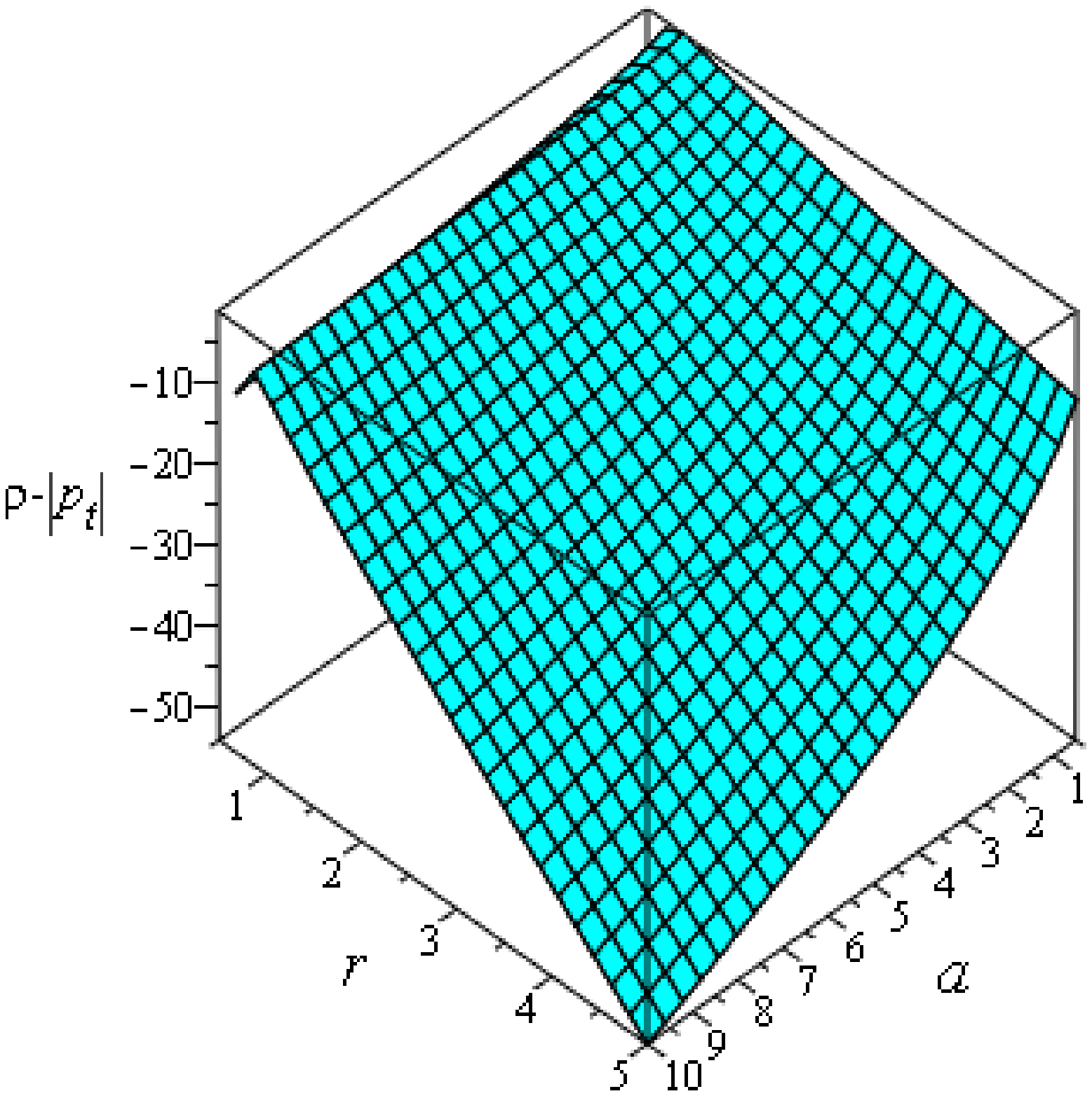}
 		\centering FIG.1(F)
 	\end{minipage}
 	\caption{Behavior of $\rho$ (FIG.1(A)), $\rho+p_r$ (FIG.1(B)), $\rho+p_t$ (FIG.1(C)), $\rho+p_r+2p_t$ (FIG.1(D)), $\rho-|p_r|$ (FIG.1(E)) and $\rho-|p_t|$ (FIG.1(F)) diagrams have been plotted  against `$r$' and `$a$' for the wormhole in model I, when $C_1=-1$ and $r_0=0.5$.}
	\label{fig3}
 \end{figure}
\begin{figure}[htb!]
	\centering
	\begin{minipage}{.55\textwidth}
		\centering
		\includegraphics[width=.6\linewidth]{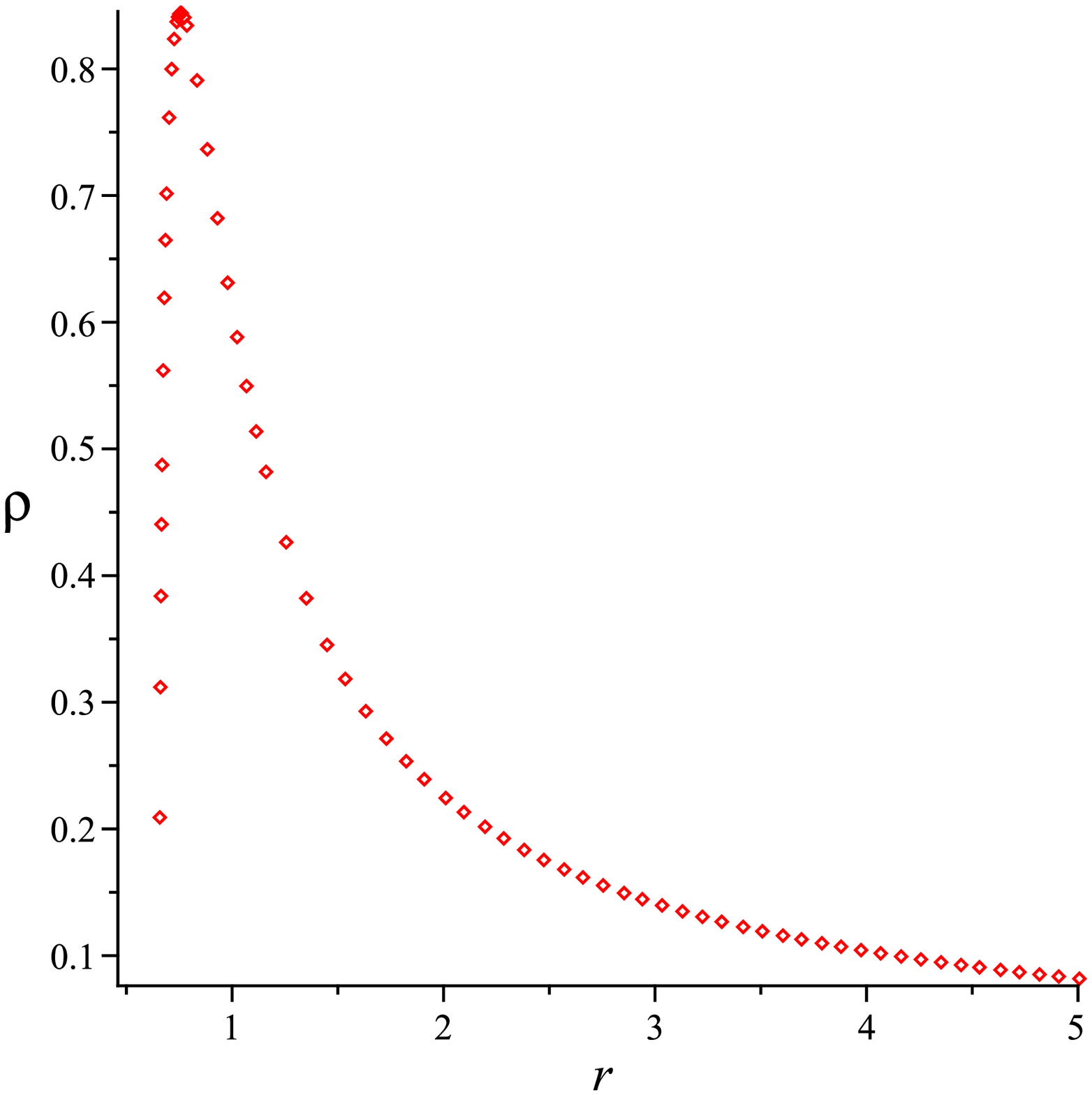}
		\centering FIG.2(A)
	\end{minipage}
	\begin{minipage}{.55\textwidth}
		\centering
		\includegraphics[width=.6\linewidth]{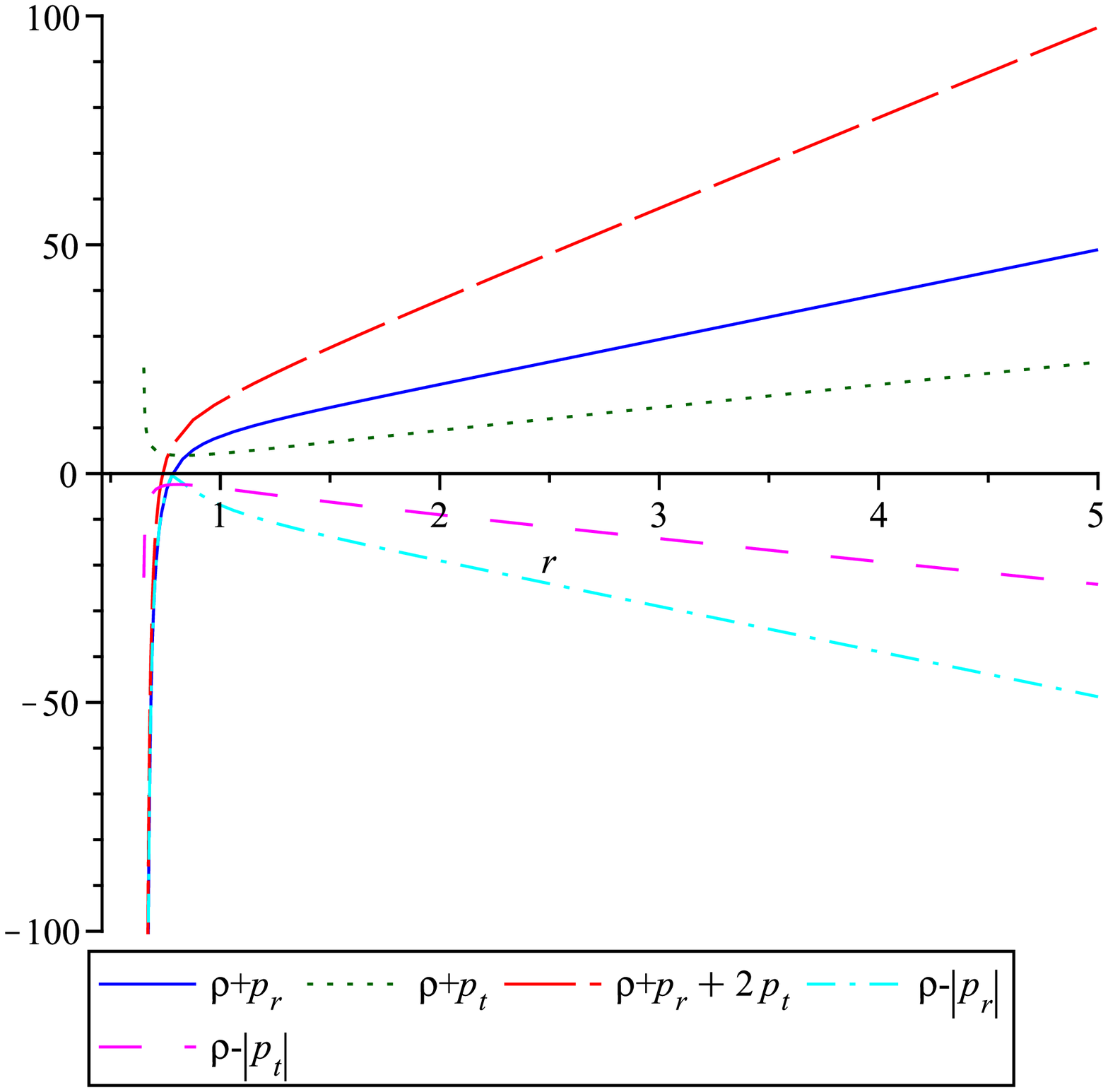}
		\centering FIG.2(B)
	\end{minipage}
	\caption{Behavior of $\rho$ (FIG.2(A)) and $\rho+p_r$, $\rho+p_t$, $\rho+p_r+2p_t$, $\rho-|p_t|$ and $\rho-|p_r|$ diagrams (FIG.2(B)) have been plotted for model I with zero tidal force against $r$ for the numerical values $a=2$, $C_1=-1$ and $r_0=0.5$.}
	\label{fig1}
\end{figure}
\begin{figure}[htb!]
	\centering
	\begin{minipage}{.42\textwidth}
		\centering
		\includegraphics[width=.6\linewidth]{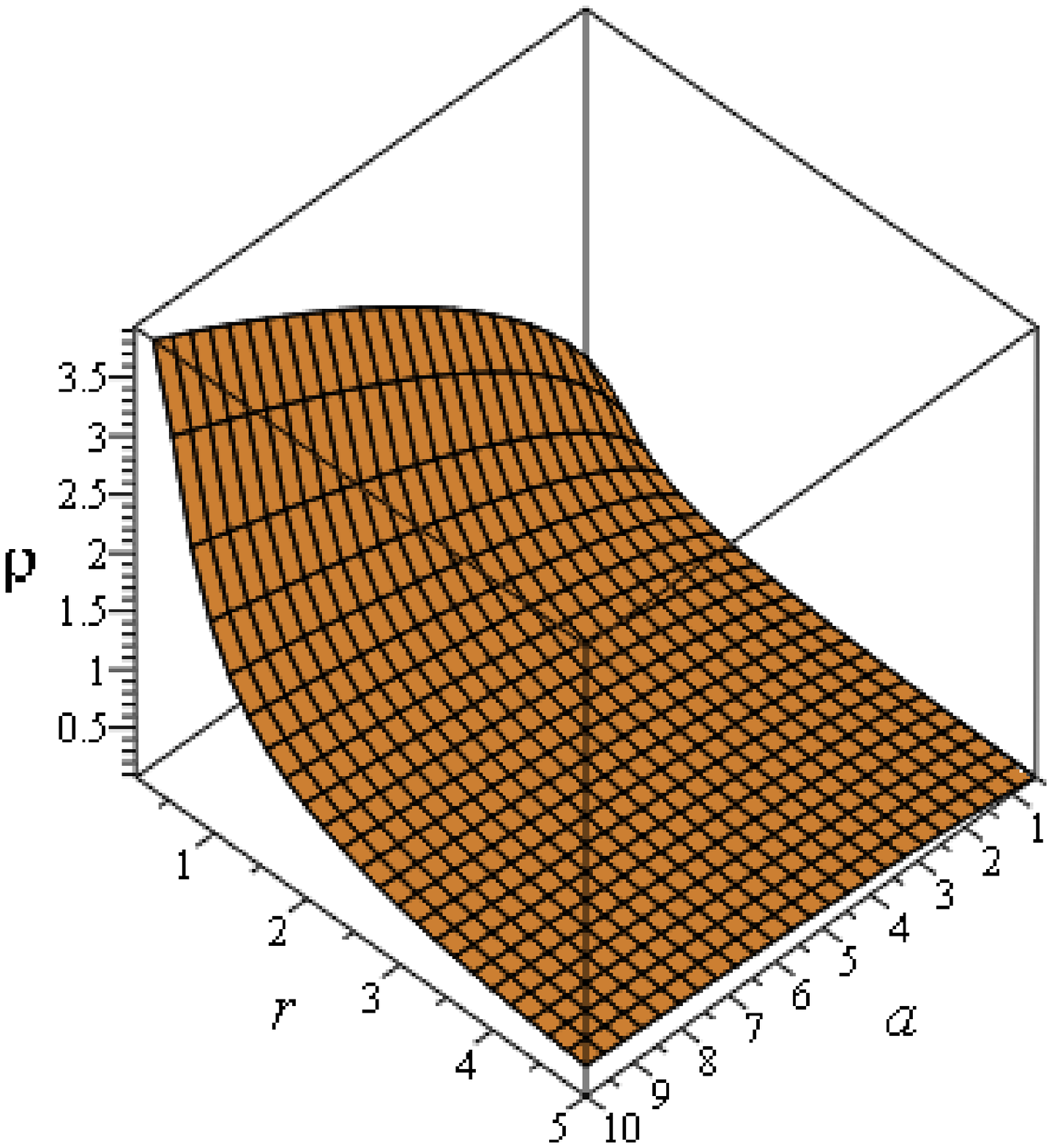}
		\centering FIG.3(A)
	\end{minipage}
	\begin{minipage}{.42\textwidth}
		\centering
		\includegraphics[width=.6\linewidth]{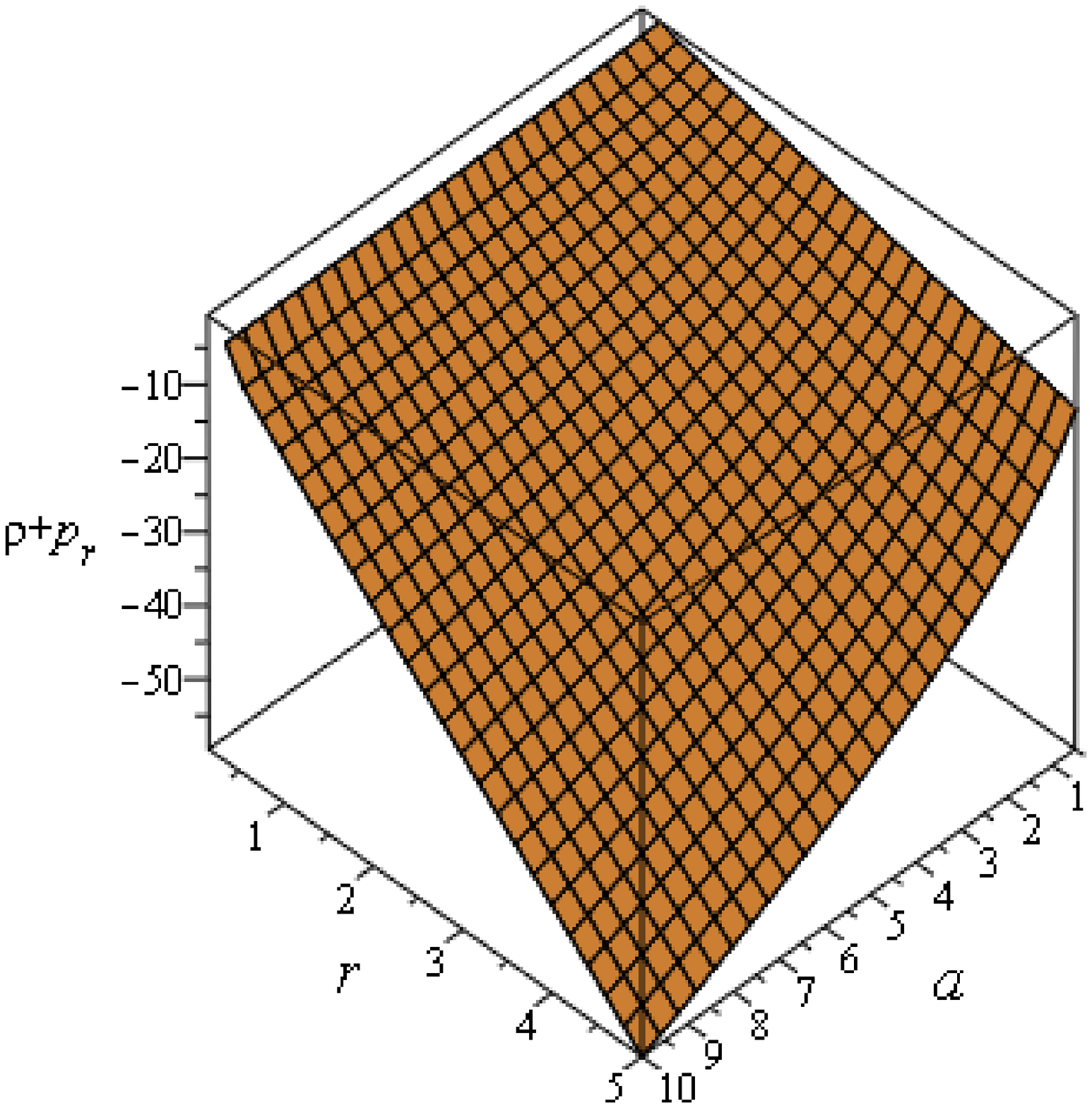}
		\centering FIG.3(B)
	\end{minipage}
  \centering
	\begin{minipage}{.42\textwidth}
		\centering
		\includegraphics[width=.6\linewidth]{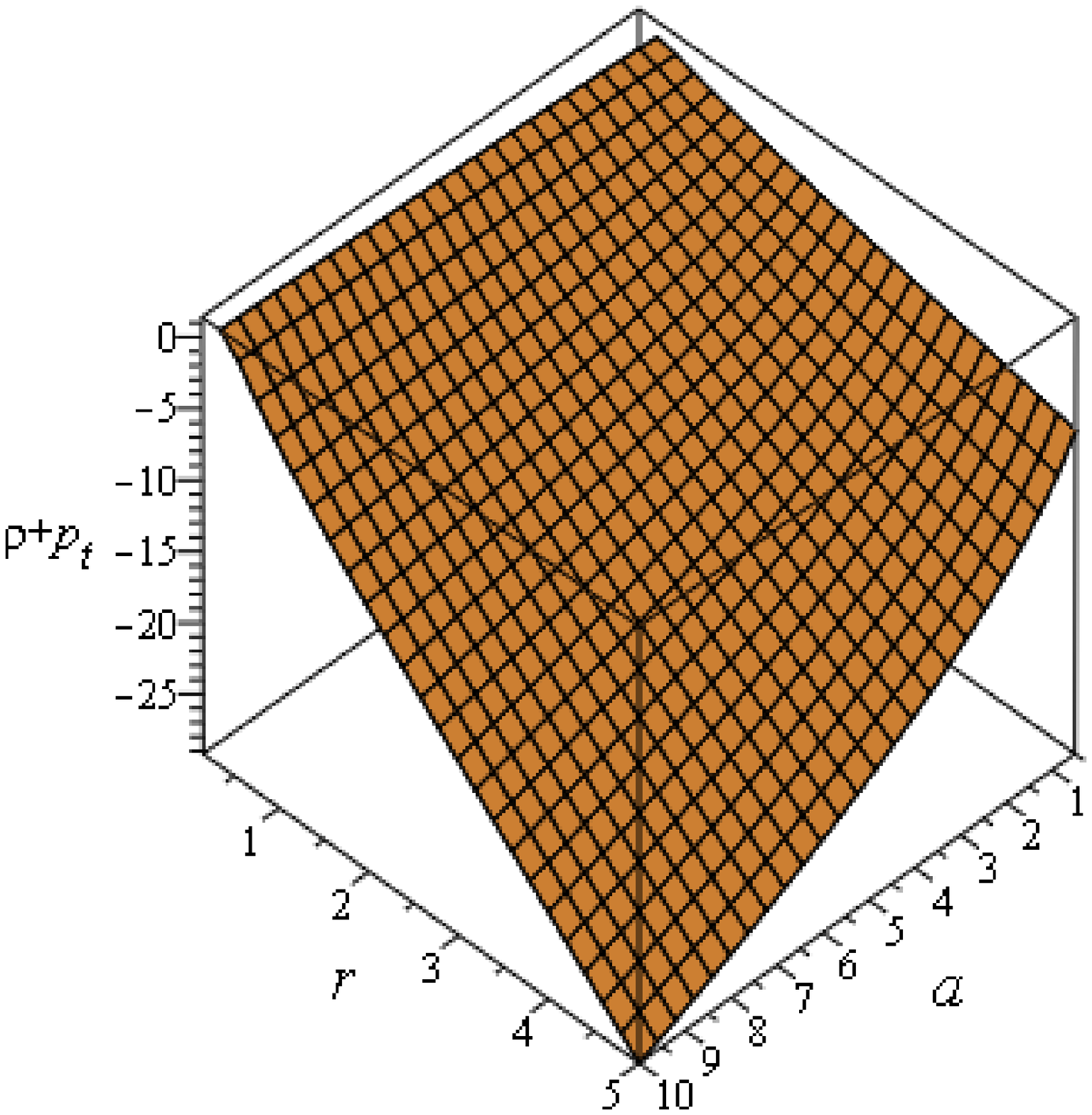}
		\centering FIG.3(C)
	\end{minipage}
	\begin{minipage}{.42\textwidth}
		\centering
		\includegraphics[width=.6\linewidth]{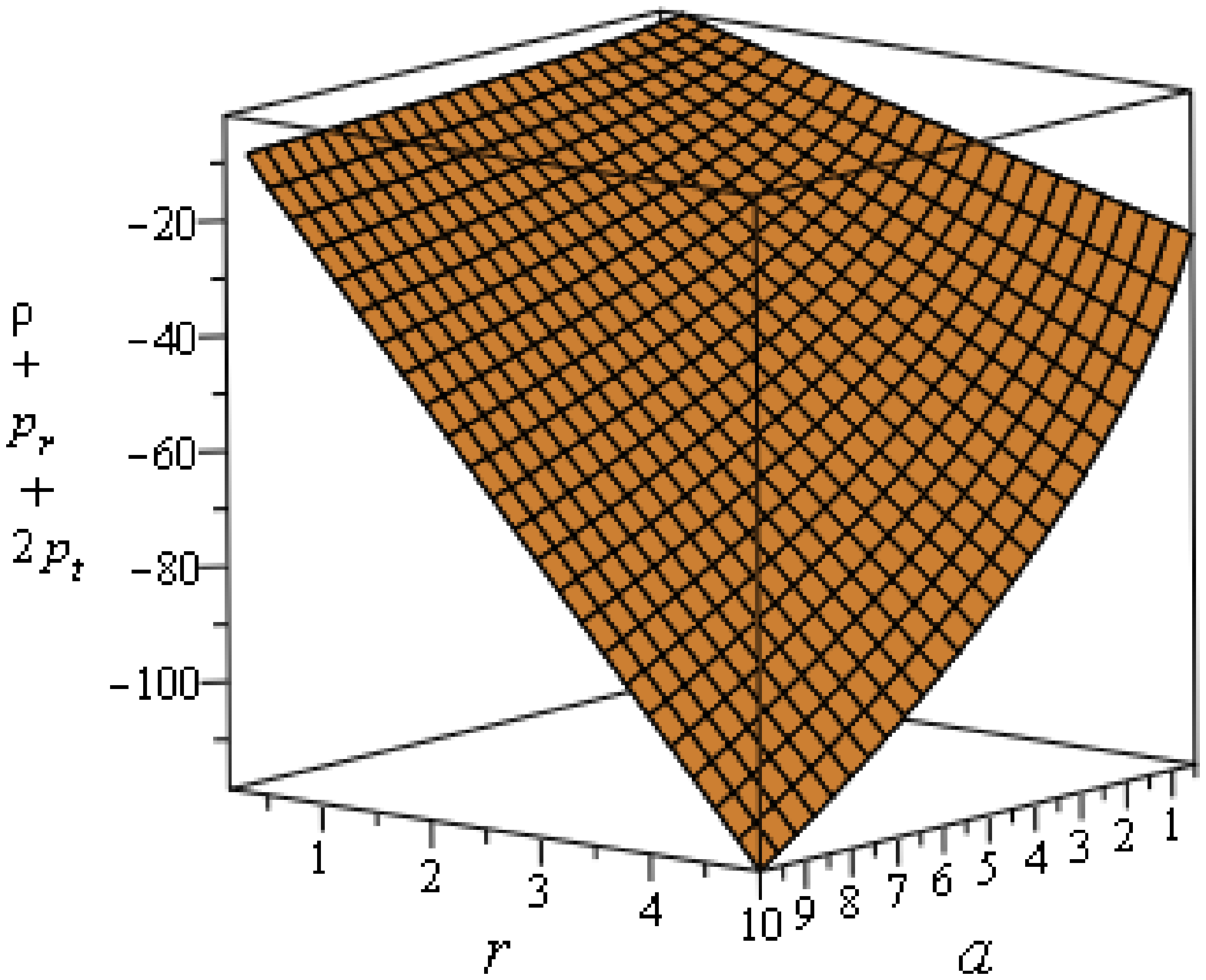}
		\centering FIG.3(D)
	\end{minipage}
  \centering
	\begin{minipage}{.42\textwidth}
		\centering
		\includegraphics[width=.6\linewidth]{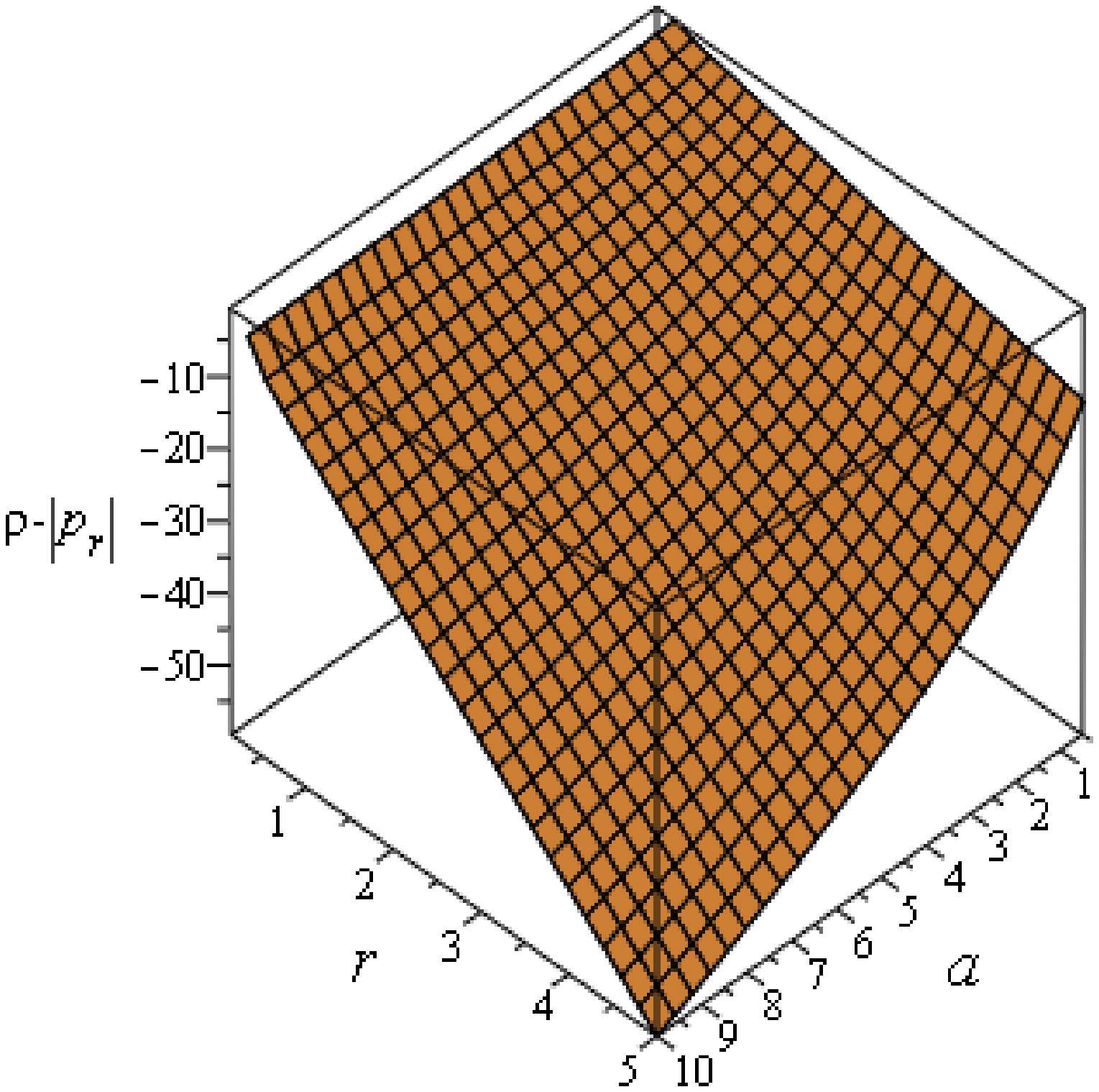}
		\centering FIG.3(E)
	\end{minipage}
	\begin{minipage}{.42\textwidth}
		\centering
		\includegraphics[width=.6\linewidth]{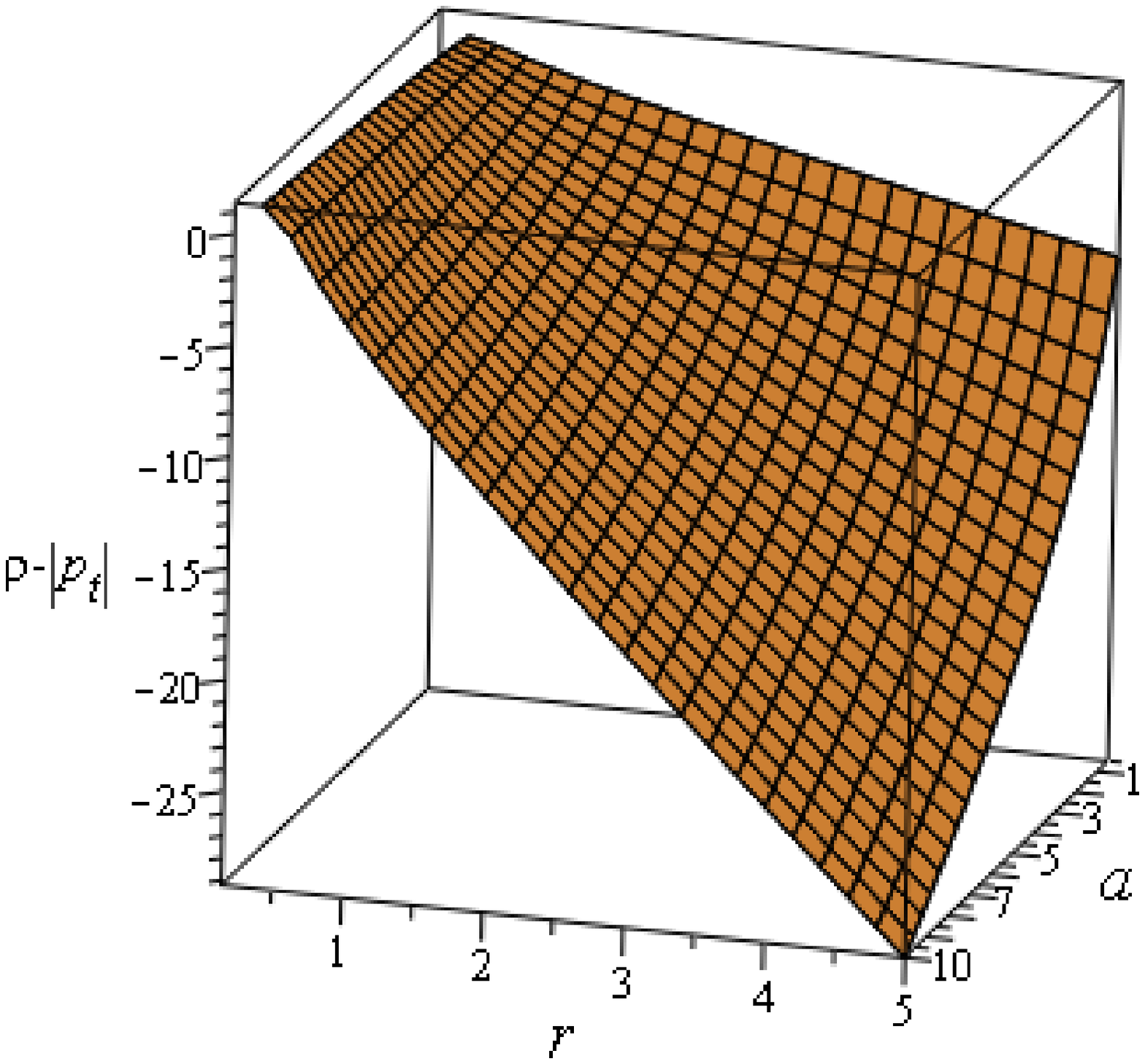}
		\centering FIG.3(F)
	\end{minipage}
	\caption{Behavior of $\rho$ (FIG.3(A)), $\rho+p_r$ (FIG.3(B)), $\rho+p_t$ (FIG.3(C)), $\rho+p_r+2p_t$ (FIG.3(D)), $\rho-|p_r|$ (FIG.3(E)) and $\rho-|p_t|$ (FIG.3(F)) diagrams have been plotted  against `$r$' and `$a$' for the wormhole in model II, when $C_2=1$ and $r_0=0.15$.}
		\label{fig3.1}
\end{figure}

\begin{figure}[htb!]
	\centering
	\begin{minipage}{.55\textwidth}
		\centering
		\includegraphics[width=.6\linewidth]{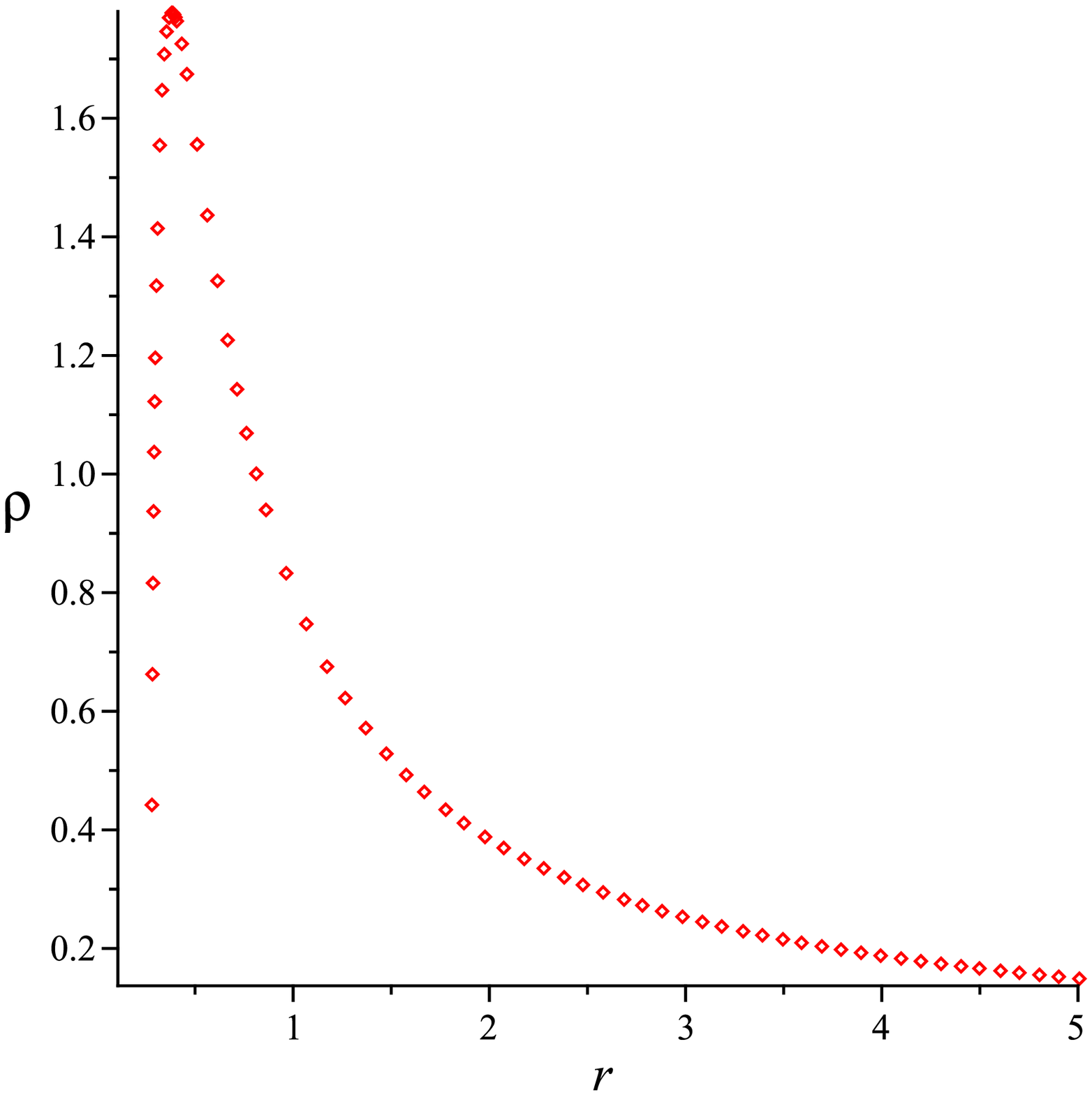}
		\centering FIG.4(A)
	\end{minipage}
	\begin{minipage}{.55\textwidth}
		\centering
		\includegraphics[width=.6\linewidth]{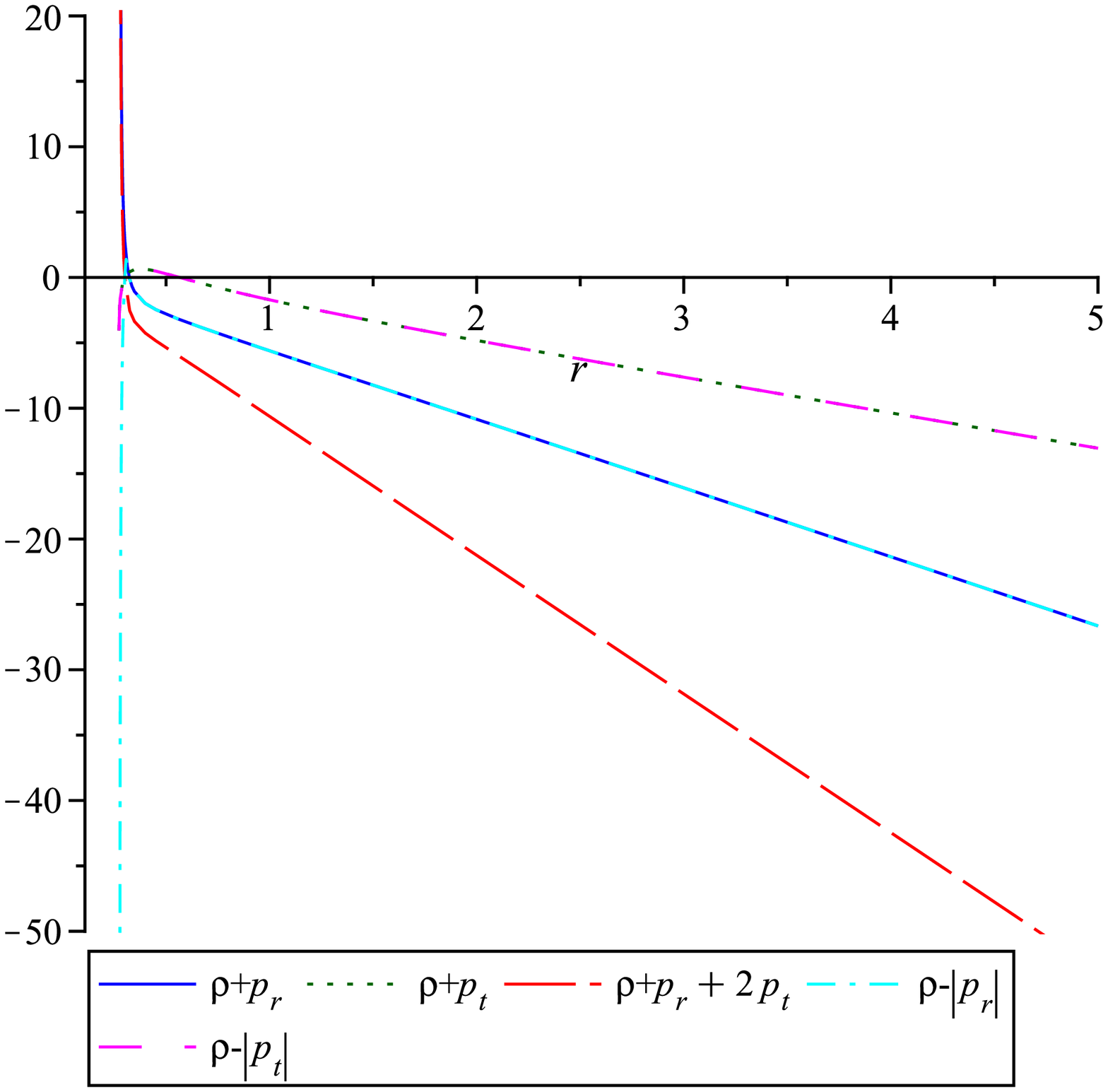}
		\centering FIG.4(B)
	\end{minipage}
	\caption{Behavior of $\rho$ (FIG.4(A)) and $\rho+p_r$, $\rho+p_t$, $\rho+p_r+2p_t$, $\rho-|p_t|$ and $\rho-|p_r|$ diagrams (FIG.4(B)) have been plotted for model II with zero tidal force against $r$ for the numerical values $a=2$, $C_2=1$ and $r_0=0.15$.}
	\label{fig7}
\end{figure}

\section{Mathematical Embedding and Proper-radial distance of wormhole}\label{sec4}
\subsection{The Embedding}
\par
\begin{figure}[htb!]
	\centering
	\begin{minipage}{.55\textwidth}
		\centering
		\includegraphics[width=.6\linewidth]{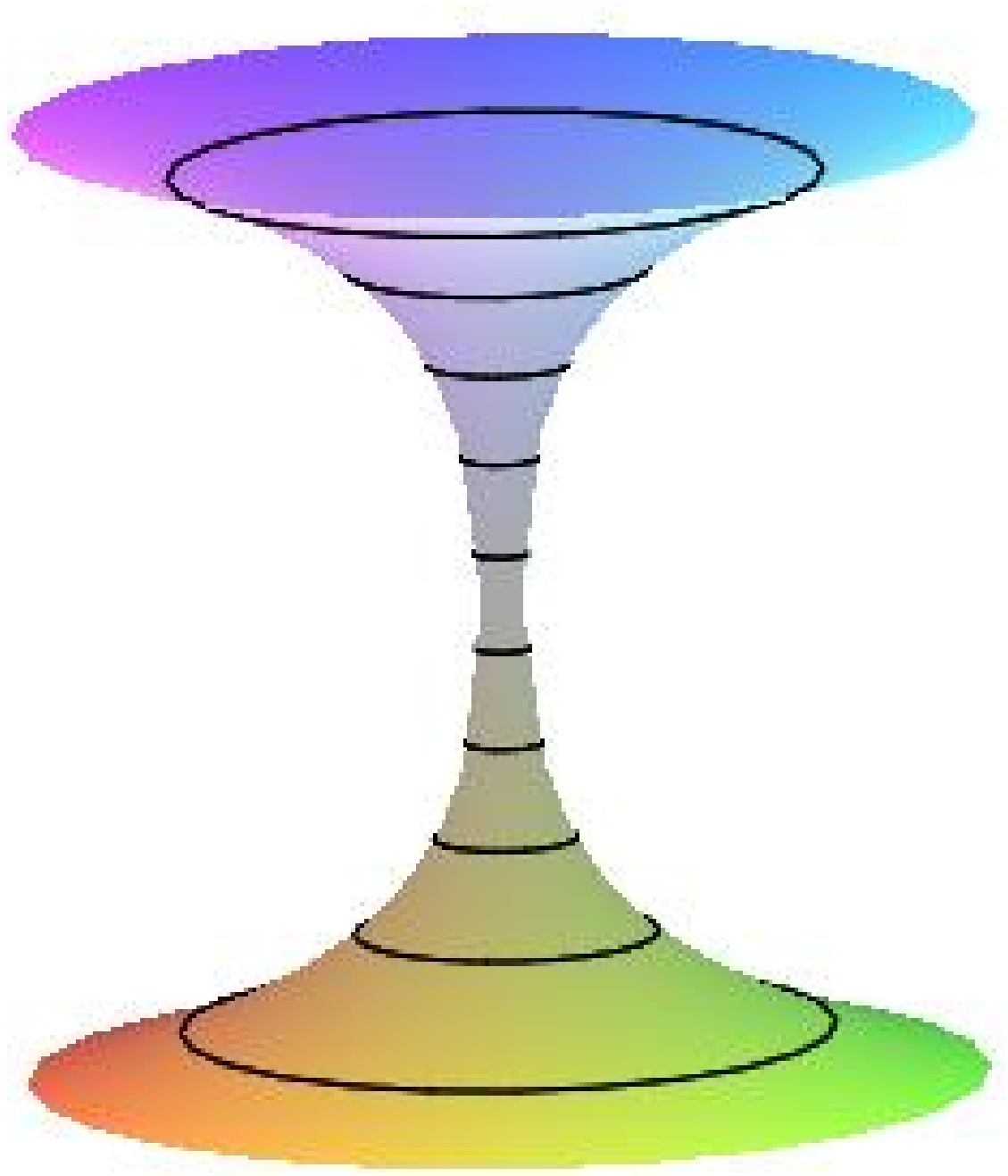}
		\centering 
	\end{minipage}
	\begin{minipage}{.55\textwidth}
		\centering
		\includegraphics[width=.6\linewidth]{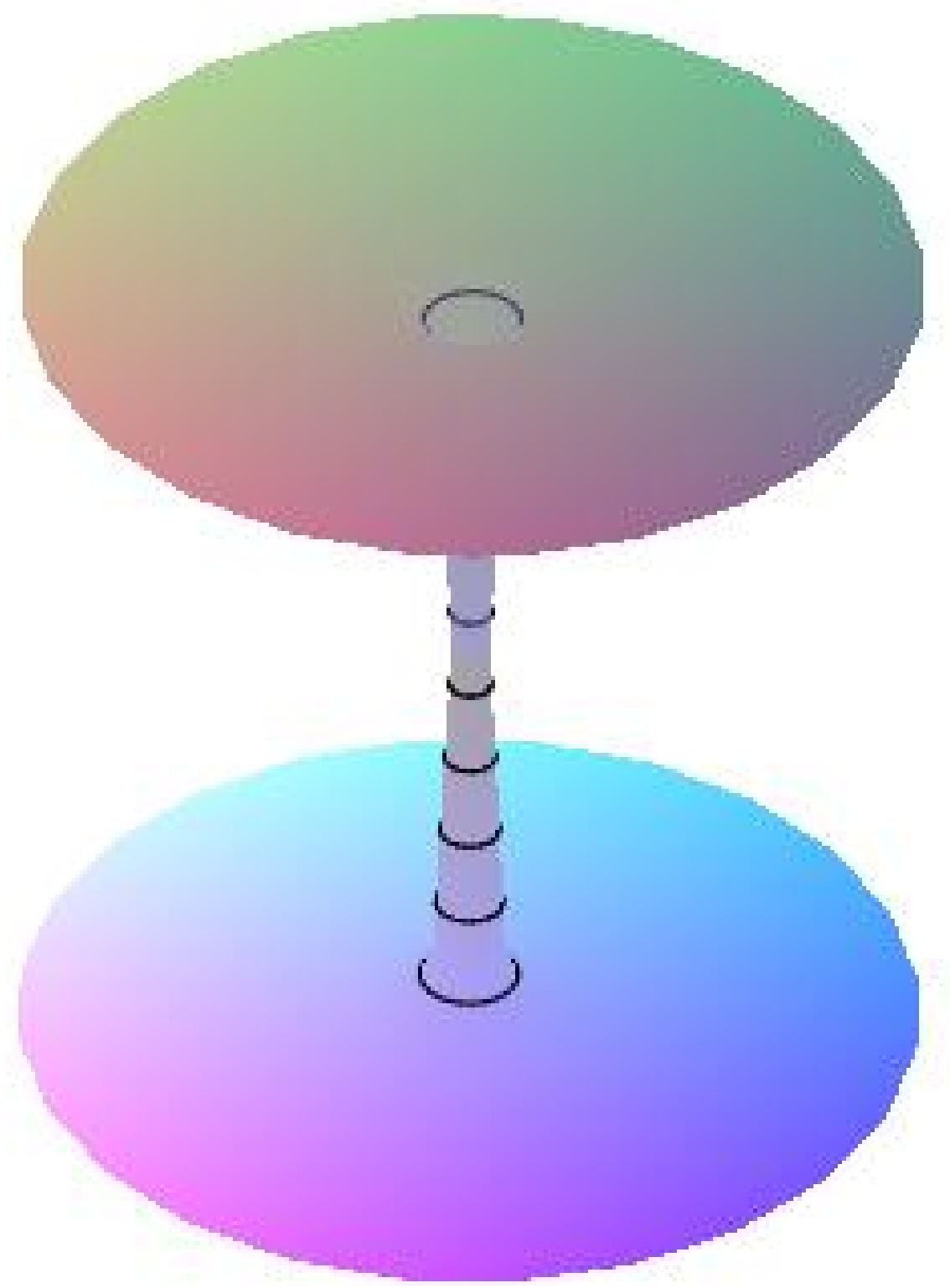}
		\centering
	\end{minipage}
	\caption{Embedding diagram for the shape function $b(r)=\frac{r_0^2}{r}$(left plot), for the shape function $b(r)=\frac{r}{1+r-r_0}$(right plot) with the numerical value $r_0=0.5$ for both shape functions. }
	\label{figemb}
\end{figure}

 Here, we consider the embedding diagram of the wormhole for two different mentioned shape functions. To visualize wormhole geometry, let us consider the two dimensional hypersurface $\tilde{W}:$ $\theta=\pi/2$, $t=$constant, without loss of generality. The respective line element is given by
\begin{equation}\label{e1}
	dS_{\tilde{W}}^2=\frac{dr^2}{1-\frac{b(r)}{r}}+r^2d\phi^2 .
\end{equation}
Now considering this hypersurface $\tilde{W}$, one embeds this metric into three dimensional Euclidean space with metric\cite{r46}
\begin{equation}\label{e2}
dS_{\tilde{W}}^2=\Big[1+\left(\frac{dz}{dr}\right)^2\Big]dr^2+r^2d\phi^2 ,
\end{equation}
as rotational surface $z=z(r,\phi)$ in cylindrical coordinates $(r,\phi,z)$. Thus, comparing equation (\ref{e1}) with equation (\ref{e2}) one deduces the equation for the embedding surface as
\begin{equation}\label{e3}
\frac{dz}{dr}=\pm\left(\frac{r}{b(r)}-1\right)^{-\frac{1}{2}} .
\end{equation}
Equation (\ref{e3}) gives the expression of $z(r)$ in the following integral form
\begin{equation}
z(r)=\pm\int_{r_0}^{r}\left(\frac{r}{b(r)}-1\right)^{-\frac{1}{2}}dr .
\end{equation}
The embedding diagrams of wormholes are drawn in figure (\ref{figemb}), considering the mentioned shape functions (which are used in models I and II).
\subsection{ Proper radial distance}
\par
\begin{figure}
	\centering
	\includegraphics[width=.5\linewidth]{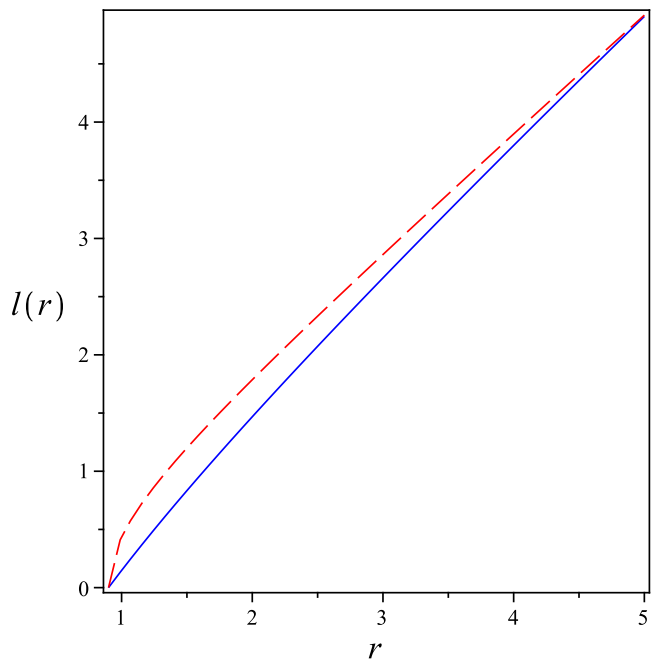}
	\caption{Behaviors of $l(r)$ with respect to the radial coordinate `$r$' corresponding to shape function $b(r)=\frac{r_0^2}{r}$(dashed line), $b(r)=\frac{r}{1+r-r_0}$(solid line) with numerical value $r_0=0.9$ for both cases.}
	\label{fig0}
\end{figure}

 In proper radial coordinate system, wormhole metric (\ref{eq1}) can be written as\cite{r46}-\cite{r47} 
\begin{equation}
ds^2=-e^{2\Phi(l)}dt^2+dl^2+r^2(l)d\Omega_2^2,
\end{equation} 
where the proper radial distance $l$ is given by 
\begin{equation}
l(r)=\int_{r_0}^{r}\frac{dr}{\sqrt{1-\frac{b(r)}{r}}} .
\end{equation}
This proper radial coordinate $l$ extends over the wormhole spacetime while the radial coordinate `$r$' ranges from $r_0$ to infinity. For both of two models, the sketches of proper radial distance $l(r)$ are shown in figure (\ref{fig0}).
\section{Results and Discussion}
\label{sec5}
\begin{table}[!]
	\centering
	\caption{Range of `$r$' where the terms of energy conditions give the result for model I and model II for $a=2$:}
	\begin{tabular}{|>{\bfseries}c|*{7}{c|}}\hline
		{\bfseries Models}  & {$\rho\geq0$} &{$\rho+p_r\geq0$} & {$\rho+p_t\geq0$} & {$\rho+p_r+2p_t\geq0$} & {$\rho-|p_r|\geq0$} &{$\rho-|p_t|\geq0$} 
		\\ \hline
		{\bfseries Model I}         &  \text{$(0.5, 5)$} &\text{$(0.8, 5)$} & \text{$(0.5, 5)$}& \text{$(0.8,5)$} &\text{Does not exist} &\text{Does not exist}\\ 
		& & & & & \text{any region}& \text{any region} \\ \hline
		{\bfseries Model II} &  \text{$(0.15,5)$} &  \text{$(0.15, 0.35)$} & \text{$(0.3,0.5)$} &\text{$(0.15, 0.25)$} &$(0.3,0.35)$&$(0.3,0.5)$
		\\ \hline
	\end{tabular}
	\label{Table:T1}
\end{table}

The wormhole metric is defined in terms of redshift and shape functions. We have considered constant redshift function throughout in this work. We have obtained two wormhole solutions for different two shape functions. Using equation (\ref{eq16}), for any differentiable shape function, the form of $f(R)$ can be obtained in the Chaplygin gas context ($p_t=-\-\frac{a}{\rho}$) but their expression may be complicated depending upon the form of $b(r)$. The energy conditions are observed to obtain the wormhole geometries in the framework of model I and model II.
 \par 
In model I, the energy conditions are examined and obeserved in the figure (\ref{fig1}) under the consideration $a=2,~C_1=-1$ and $r_0=0.5$.\\ In this case, the energy density is positive for $r_0=0.5<r<5$. Both terms of DEC are negative for all $r\in(r_0, 5)$. Hence, from table(\ref{Table:T1}) we have found a region $0.8<r<5$ where all energy conditions are satisfied except DEC.
   \par 
    In model II, the energy conditions are examined and obeserved in the figure (\ref{fig7}) under the consideration $a=2,~C_2=1$ and $r_0=0.15$.\\ 
   Also in this case, the energy density is non-negative every where {\it i.e.} for all $r\in(r_0=0.15, 5)$. Here NEC and WEC are satisfied for $r\in(0.3,0.35)$. The SEC is not satisfied in any region. And DEC is satisfied for $r\in(0.3,0.35)$(see table \ref{Table:T1}).
   \par 
   From the above, it is seen that we have found a region where all the energy conditions are satisfied in some cases. But in the neighbourhood of the wormhole throat none of the energy condition is satisfied in any cases. From figure (\ref{fig3}) and figure (\ref{fig3.1}), it is observed that all energy conditions are not valid in the neighbourhood of wormhole throat for $0.5\leq a\leq10$, however it also observed that a major energy conditions are satisfied in a region. Thus the presence of exotic matter is necessary to form of wormhole in this scenario.

	\end{document}